\documentclass[10pt]{article}
\usepackage{hyperref}
\usepackage{graphicx, subfigure}

\newcommand{\fb}{$\mathrm{fb}^{-1}$}
\newcommand{\sigmaSM}{$\sigma_\mathrm{SM}$}
\newcommand{\sigmaBR}{$\sigma\times{BR}$}
\newcommand{\gammad}{$\gamma_\mathrm{d}$}
\newcommand{\higgstwo}{\mbox{Higgs $\rightarrow$ 2 \gammad}}
\newcommand{\higgsfour}{\mbox{Higgs $\rightarrow$ 4 \gammad}}
\newcommand{\mH}{$m_\mathrm{H}$}
\newcommand{\pt}{$p_\mathrm{T}$}

\def\Title#1{\begin{center} {\Large #1 } \end{center}}
\def\Author#1{\begin{center}{ \sc #1} \end{center}}
\def\Address#1{\begin{center}{ \it #1} \end{center}}

\newcommand\pubblock{\rightline{\begin{tabular}{l} Proceedings of the Fifth Annual LHCP\\ \pubnumber\\
         \pubdate  \end{tabular}}}

\newenvironment{Abstract}{\begin{quotation} \begin{center} 
             \large ABSTRACT \end{center}\bigskip 
      \begin{center}\begin{large}}{\end{large}\end{center} \end{quotation}}

\newenvironment{Presented}{\begin{quotation} \begin{center} 
             PRESENTED AT\end{center}\bigskip 
      \begin{center}\begin{large}}{\end{large}\end{center} \end{quotation}}

\def\beq{\begin{equation}}
\def\eeq#1{\label{#1}\end{equation}}
\def\eeqn{\end{equation}}

\def\beqa{\begin{eqnarray}}
\def\eeqa#1{\label{#1}\end{eqnarray}}
\def\eeqan{\end{eqnarray}}

\let\bar=\overbar

\def\Dslash{\not{\hbox{\kern-4pt $D$}}}
\def\dslash{\not{\hbox{\kern-2pt $\del$}}}

\def\msb{{\bar{\ssstyle M \kern -1pt S}}}

\usepackage{color}

\newcommand\pubnumber{ ATL-PHYS-PROC-2017-098 }

\newcommand\pubdate{\today}

\def\affiliation{
On behalf of the ATLAS Collaboration, \\
Department of Physics \\
University of Calabria and INFN-Cosenza, Arcavacata di Rende, 87036, Italy}

\begin{document}

\large
\begin{titlepage}
\pubblock

\vfill
\Title{Search for long-lived neutral particles
decaying into Lepton-Jets with the ATLAS detector
in proton-proton collision data at 13 TeV}
\vfill

\Author{ DANIELA SALVATORE  }
\Address{\affiliation}
\vfill
\begin{Abstract}

Several models of particle physics different from the Standard Model predict the existence of a dark sector that is
weakly coupled to the visible one: the two sectors may couple via the vector portal, where a dark
photon with mass in the MeV to GeV range mixes kinetically with the SM photon. If the dark
photon is the lightest state in the dark sector, it will decay to SM particles, mainly to leptons and possibly light mesons. Due to its weak interactions with the SM, it can have a non-negligible lifetime. At the LHC, these dark photons would typically be produced with large boost resulting in collimated jet-like structures containing pairs of leptons and/or light hadrons, the so-called Lepton-Jets.

This work focuses on the search for displaced Lepton-Jets, which are produced away from the interaction point and their constituents are limited to electrons, muons, and pions. 
Results from the anlaysis of 3.4~\fb proton-proton collision data sample recorded by ATLAS at a center-of-mass energy of 13~TeV during 2015 are compared to the Standard Model expectations and with BSM predictions.

\end{Abstract}
\vfill

\begin{Presented}
The Fifth Annual Conference\\
 on Large Hadron Collider Physics \\
Shanghai Jiao Tong University, Shanghai, China\\
May 15-20, 2017
\end{Presented}
\vfill
\end{titlepage}
\def\thefootnote{\fnsymbol{footnote}}
\setcounter{footnote}{0}
%

\normalsize 


\section{Introduction}

Many possible extensions of the Standard Model (SM) predict a dark sector that is weakly coupled to the visible one~\cite{Strass}.
Depending on the structure of the dark sector and its
coupling to the SM, some unstable dark states may be produced at colliders and decay back to SM
particles with sizeable branching fractions. An extensively studied case is one in which the two
sectors couple via the vector portal, where a dark photon (\gammad) with mass in the MeV to GeV range
mixes kinetically with the SM photon~\cite{FRVZ}. If the \gammad~is the lightest state in the dark sector, it will
decay to SM particles, mainly to leptons and possibly light mesons. Due to its weak interactions
with the SM, it can have a non-negligible lifetime. At the LHC, these \gammad~would typically
be produced with large boost resulting in collimated jet-like structures containing pairs of leptons
and/or light hadrons (lepton-jets, LJs). If produced away from the interaction point (IP), they are
referred to as Òdisplaced LJsÓ. The search described in the present contribution looks for events
with a pair of displaced LJs originating from the decay of an heavy particle and the LJ constituents
are electrons, muons, and pions. 

\begin{figure}[htb]
\centering
\subfigure {\includegraphics[height=2in]{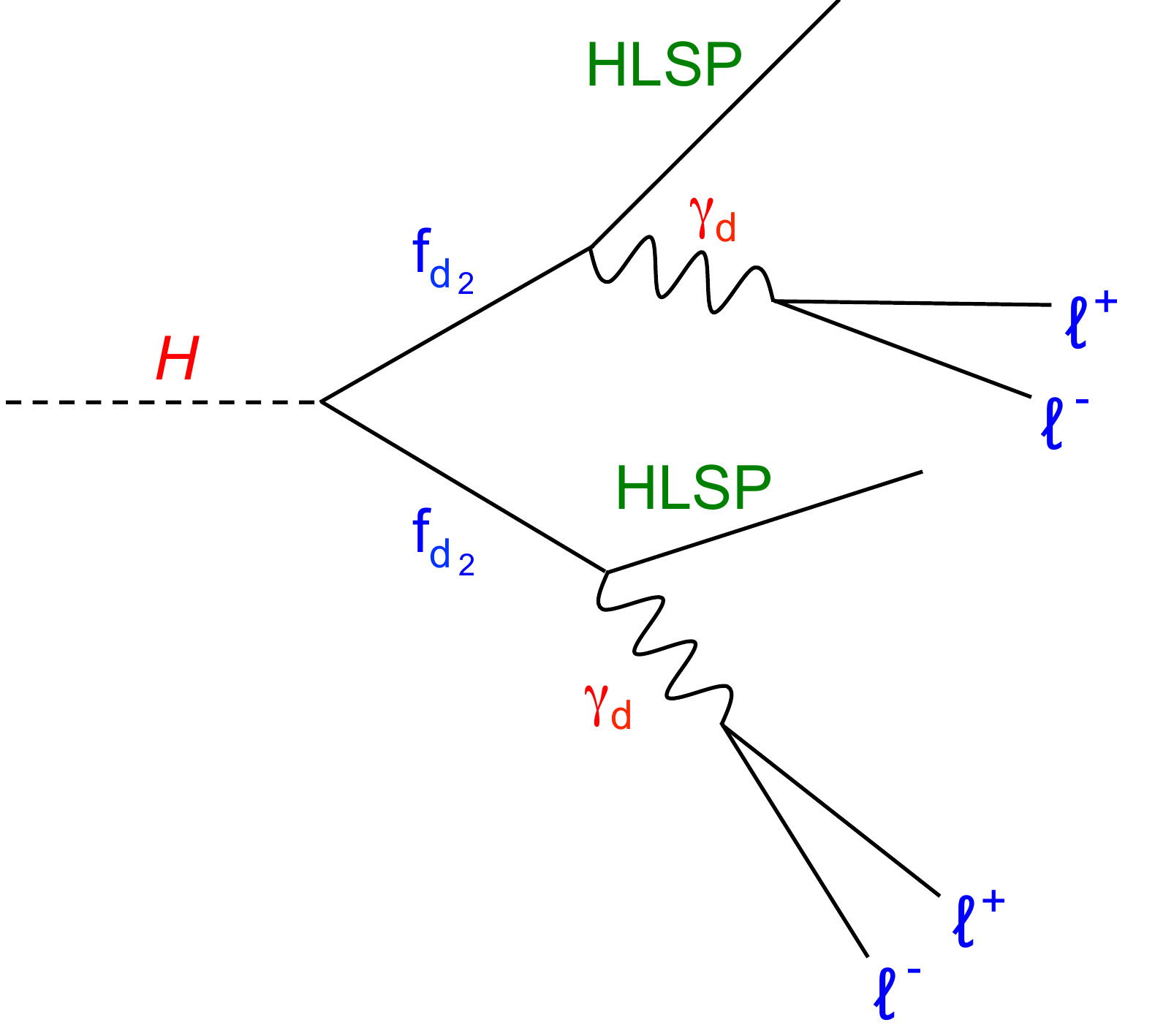}}
\subfigure {\includegraphics[height=2in]{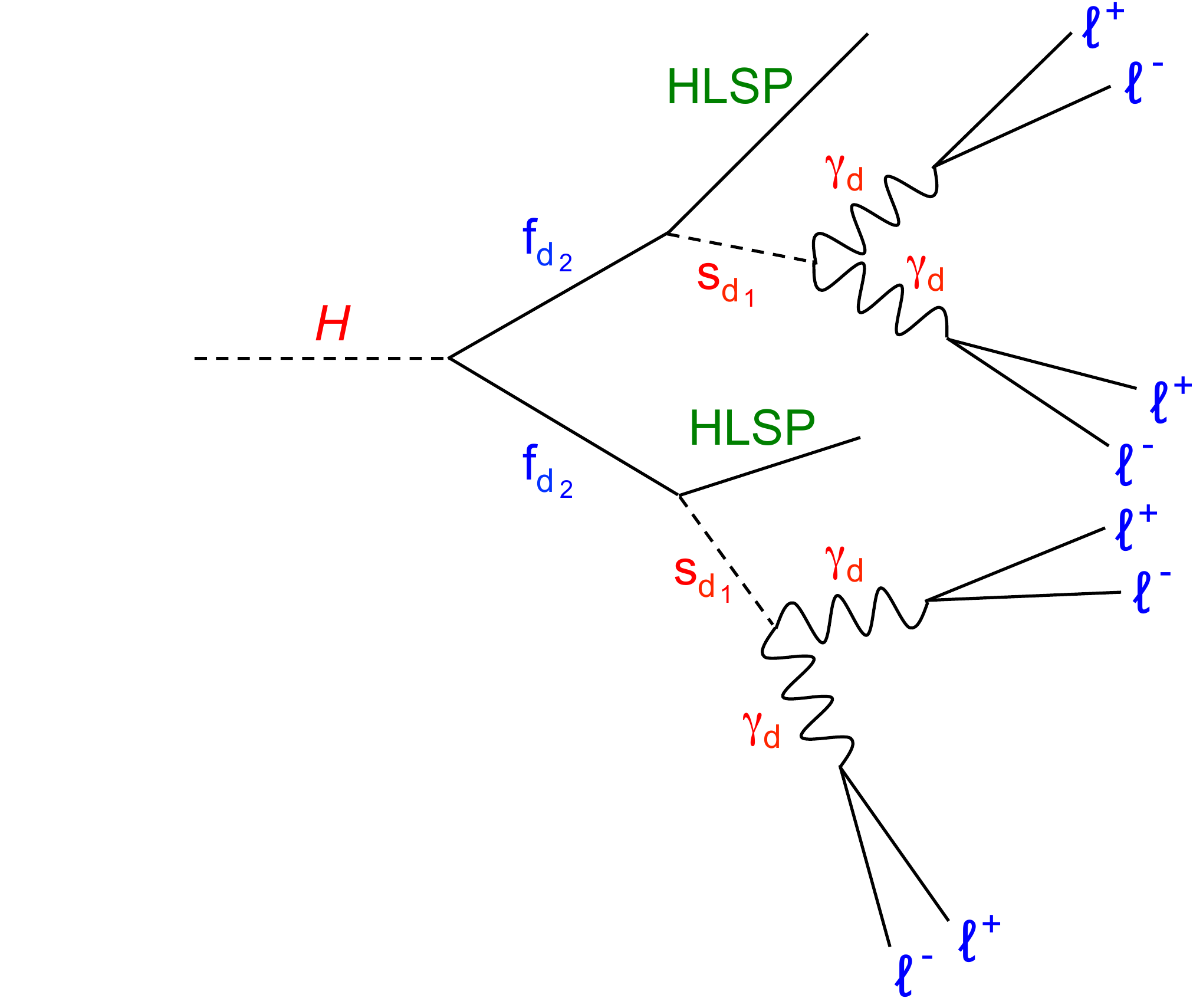}}
\caption{The two FRVZ models used as benchmarks in the analysis~\cite{LJ}.}
\label{fig:models}
\end{figure}

Results are interpreted in the context of the two Falkowski-Ruderman-Volansky-Zupan (FRVZ) models~\cite{FRVZ} which
predict non-SM Higgs boson decays to LJs. As shown in Figure~\ref{fig:models}, the first FRVZ model produces
two \gammad~while the second produces four \gammad. In the first model (left), the dark fermion decays
to a $\gamma_{d}$ and a lighter dark fermion $f_{d1}$, assumed to be the massless HLSP (Hidden Lightest Stable Particle).
In the second model (right), the dark fermion $f_{d2}$ decays to an HLSP and a dark scalar $s_{d1}$ that in
turn decays to pairs of \gammad. The mass of the \gammad~is assumed to be 400 MeV. Details
on the models are in Ref.~\cite{LJ}. The $\gamma_{d}$ decay lifetime $\tau$ (expressed as $\tau$ times the speed of light $c$) is
a free parameter of the model. 

The search employs the full dataset collected by ATLAS~\cite{Aad:2012tfa}
during 2015 at a center-of-mass energy of 13~TeV, corresponding to an integrated luminosity of 3.4~\fb~\cite{LJ}.

\section{Lepton-Jet reconstruction and selection}

\begin{figure}[!hb]
\centering
\subfigure {\includegraphics[height=2in]{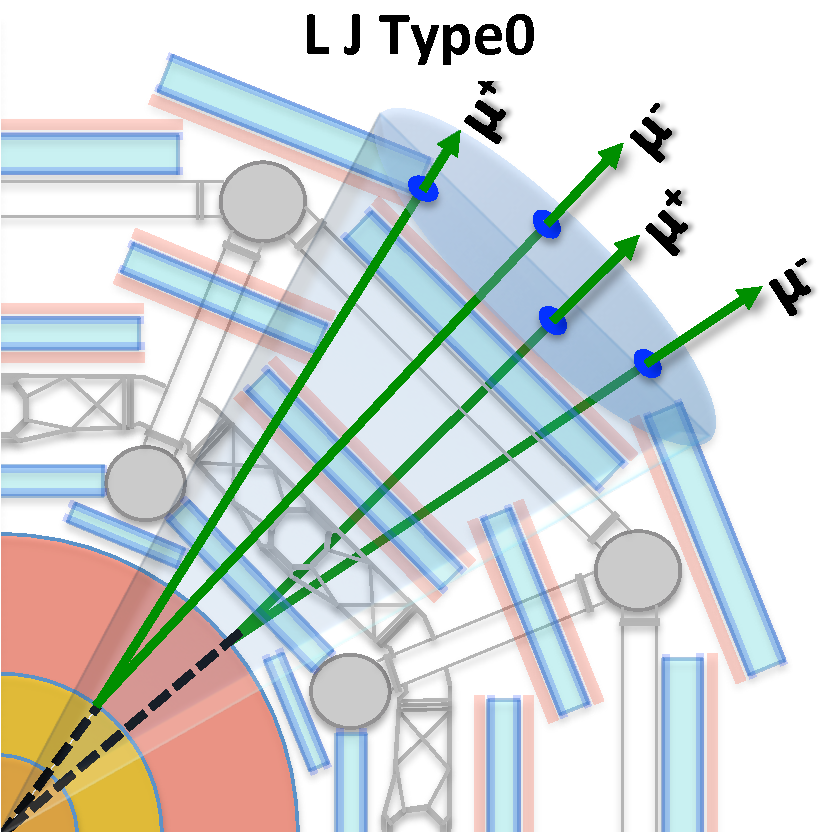}}
\subfigure {\includegraphics[height=2in]{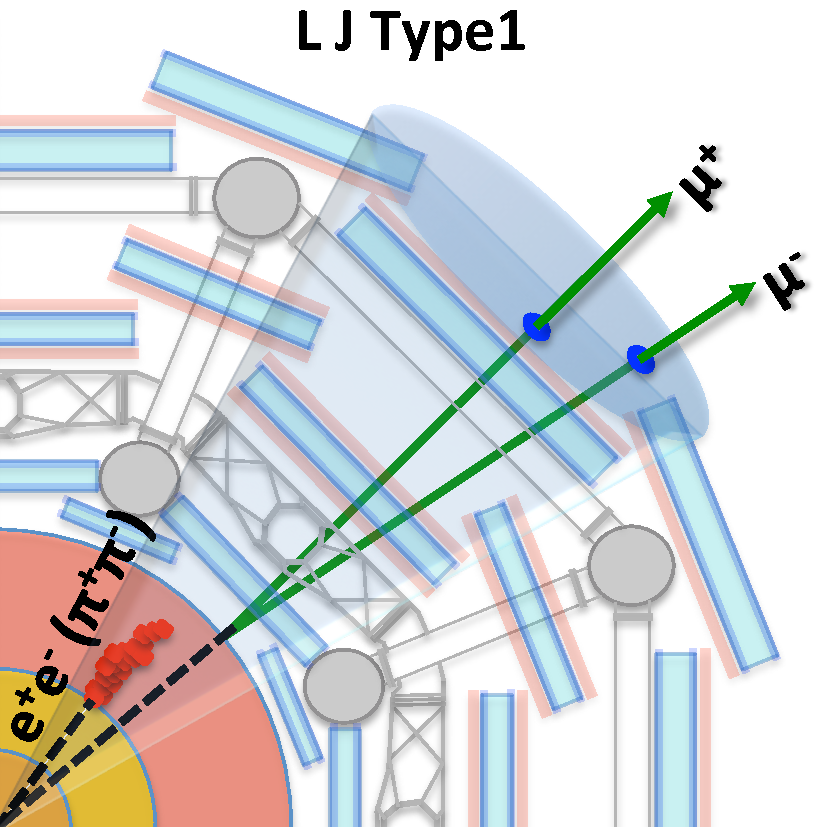}}
\subfigure {\includegraphics[height=2in]{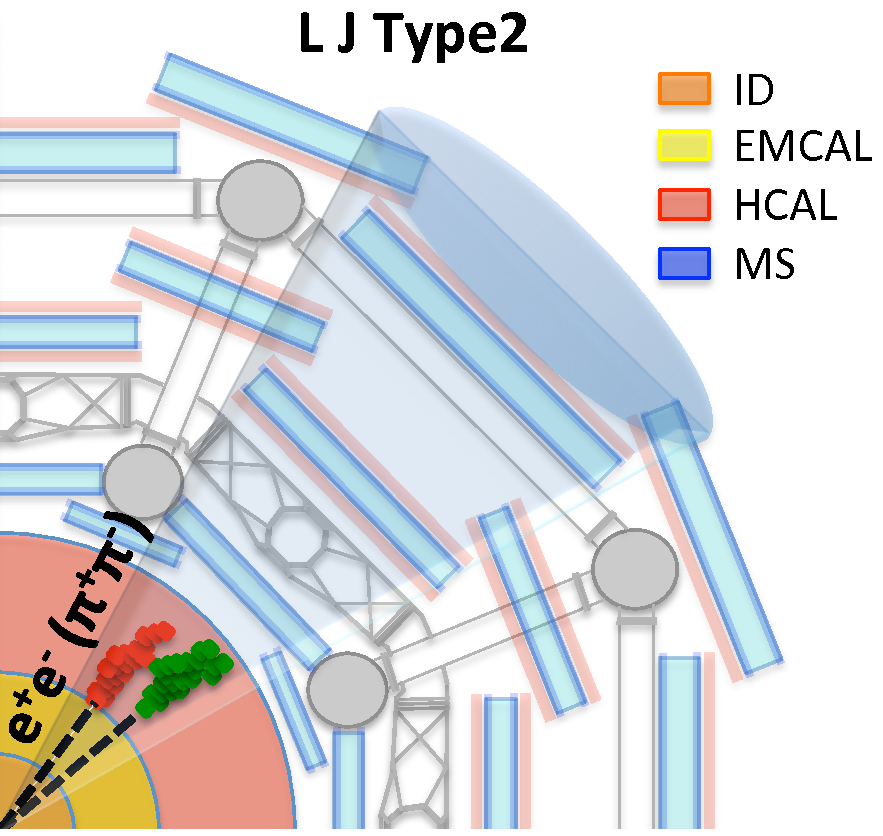}}
\caption{The three types of Lepton-Jets~\cite{LJ}.}
\label{fig:types}
\end{figure}

The search is based on a generic definition of LJ in order to make the analysis as model-independent
as possible. The LJ contains at least one \gammad~decaying far from the primary vertex, so they are expected highly isolated in the inner tracker. Three topologies of LJs are considered (Figure~\ref{fig:types}).

TYPE0 LJs are reconstructed using a simple clustering algorithm that combines all the muons within
a cone of  $\Delta R = 0.5$, the maximum
opening angle between the muons or between muons and jets in the benchmark model MC. The algorithm is seeded by the highest-\pt~muon. 
If at least two muons are found in the cone, the LJ is accepted. The search is repeated with any unassociated muon until no muon
seed is left.

A TYPE1 LJ is a cluster of at least two muons and jets in a cone opening
$\Delta R = 0.5$: it is the signature of at least one \gammad~decaying to a muon pair and at least another one
decaying to an electron or pion pair. This can occur in the FRVZ models with 4 \gammad, where the pair from the same $f_{d2}$ is very collimeated.

A TYPE2 LJ is a jet with low electromagnetic fraction and no
muons in a $\Delta R = 0.5$ cone. It is the signature of LJs produced by a single \gammad~or LJs containing two \gammad, each decaying to an electron/pion pair, always expected to be
reconstructed as a single jet due to their large boost.

The triggers used to select displaced LJs of TYPE0 and TYPE1 are multi-muon triggers with
muons reconstructed only in the muon spectrometer. For the TYPE2 LJs the single jet trigger with
low electromagnetic fraction is used: only LJs produced in the hadronic calorimeters are selected,
reducing the background due to the QCD jets. The events that pass the trigger filters have to satisfy
additional requirements to separate the signal from the two major sources of background: muons from cosmic rays and QCD multi-jet events. The cosmic-ray events
are reduced using jet timing, and for muons there is an additional requirement: the perigee to the
beam line of the muon spectrometer track has to be close to the primary vertex of the event. QCD
multi-jet background is reduced using track isolation in the inner tracker, jet electromagnetic fraction
and width of the jets. Two additional requirements are used: exactly two reconstructed LJs in
the event; the absolute value of the azimuthal angle $\Delta \phi$ between the two LJs must be $\geq 1$ (in the
geometric limits of the LJ reconstruction cone). The residual QCD multi-jet background and cosmics
background is evaluated using a data-driven matrix ABCD method. Details on the selection and background estimation can be found in Ref.~\cite{LJ}.

\section{Results}

Table~\ref{tab:res} shows the final result: the observed data events and the expected
multijet and cosmic-ray residual contaminations in the signal region. Taking into account the expected high background
for the TYPE2 - TYPE2 events, the ABCD method is performed on three categories: all
events, all events but excluding the TYPE2 - TYPE2 pairs and events containing only the TYPE2 - TYPE2 combination. In all categories,
no evidence of signal is observed.

\begin{table}[h]
\begin{center}
\begin{tabular}{l|c|c}  
Category & Observed events &  Expected background\\ \hline
All events                          & 285 & 231$\pm$ 12 (stat) $\pm$ 62 (syst)\\
TYPE2 - TYPE2 excluded & 46    & 31.8$\pm$ 3.8 (stat) $\pm$ 8.6 (syst)\\
TYPE2 - TYPE2 only         & 239 & 241$\pm$ 41(stat) $\pm$6 5(syst)\\
\end{tabular}
\caption{Results of the ABCD method compared with the observed events on data~\cite{LJ}.}
\label{tab:res}
\end{center}
\end{table}

\begin{figure}[h]
\centering
\subfigure {\includegraphics[height=2in]{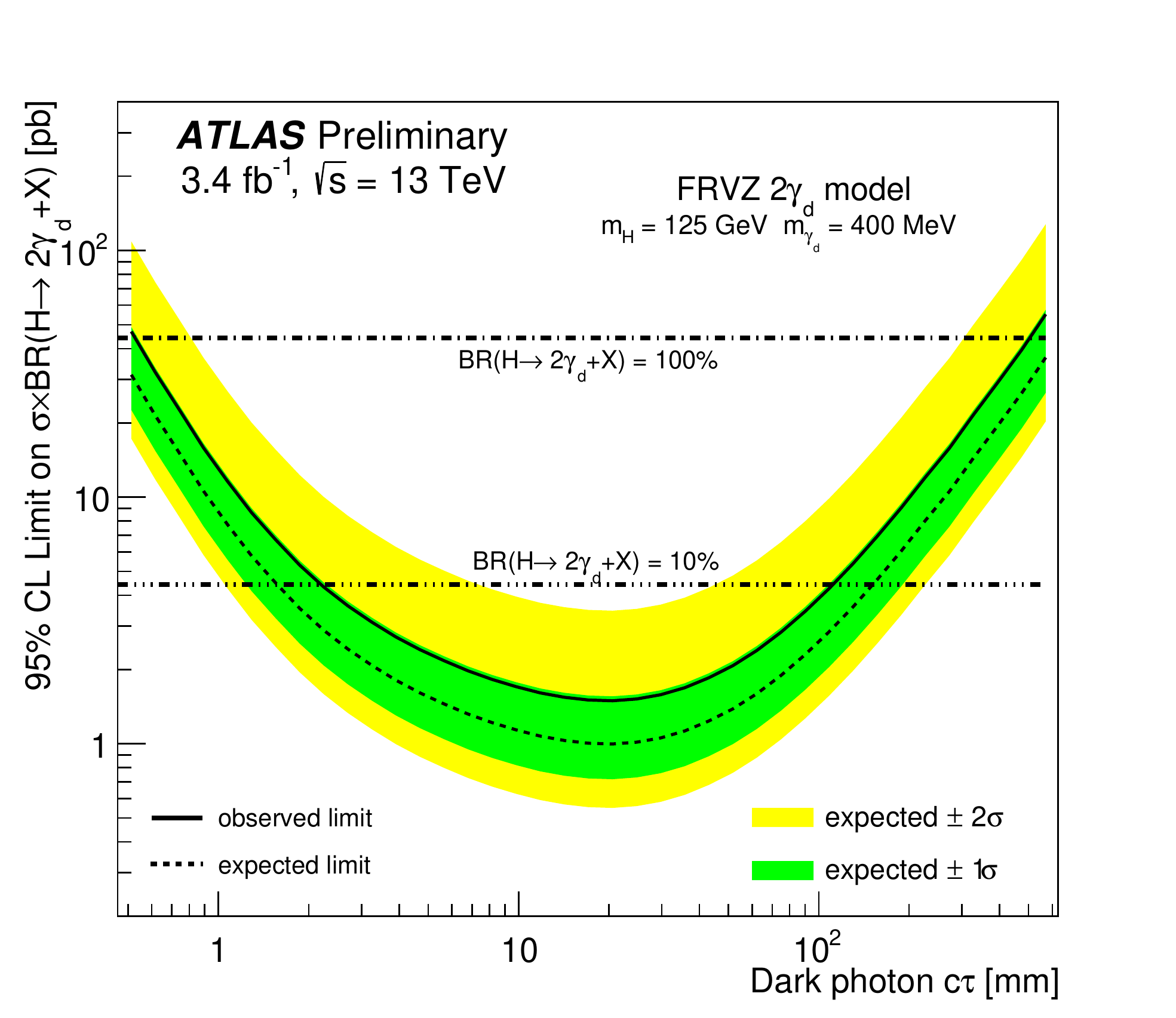}}
\subfigure {\includegraphics[height=2in]{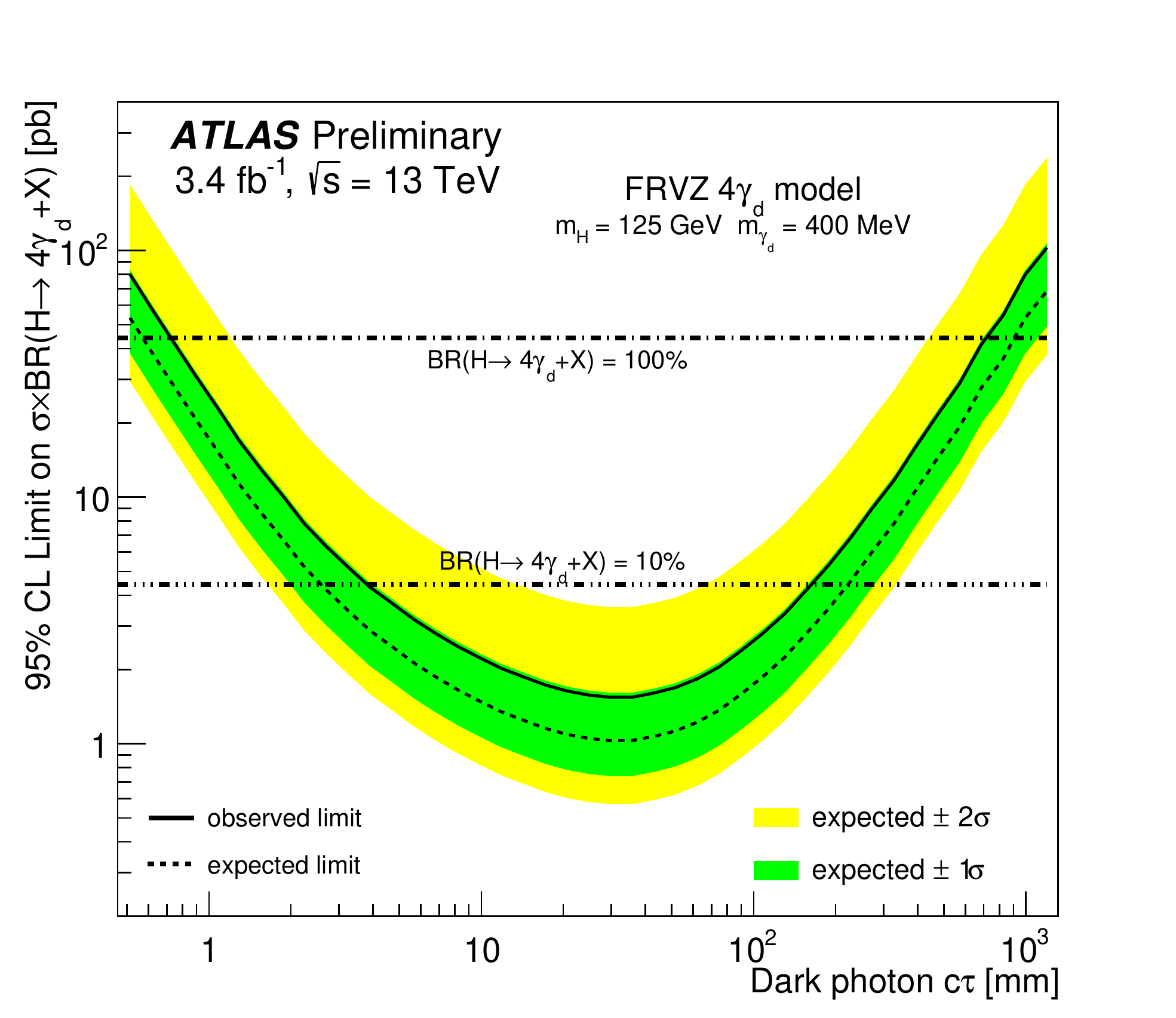}}
\caption{The 95\% upper limits on the  \sigmaBR~for the FRVZ 125 GeV Higgs \higgstwo+X (left) and \higgsfour+X (right) benchmark models as a function 
of the \gammad~lifetime ($c\tau$)~\cite{LJ}. The horizontal lines
correspond to \sigmaBR for two values of the BR of the Higgs boson decay to \gammad. TYPE2 - TYPE2
events are excluded.}
\label{fig:limits}
\end{figure}

In the absence of a signal, the results of the search for LJ production are used to set upper
limits on the product of cross section and Higgs decay branching fraction to LJs, as a function of
the $\gamma_{d}$ mean lifetime in the two FRVZ models. The $CL_{s}$ method is used to determine the exclusion limits, where
the signal region is populated from the data-driven background estimate and from the appropriate
signal hypothesis. The resulting exclusion limits on the \sigmaBR, assuming the gluon-fusion production cross section of the 125 GeV SM Higgs boson \sigmaSM = 44.13 pb are shown in Figure~\ref{fig:limits} as a function of
the \gammad~mean lifetime for the \higgstwo+X and \higgsfour+X models. 
Taking into account the relatively
low signal efficiency for the TYPE2 - TYPE2 event in all models and the high background, the TYPE2 - TYPE2
events are excluded from the limit evaluation. The expected limit is shown as the dashed curve and
the solid curve shows the observed limit. The horizontal lines correspond to \sigmaBR~ for two values
of the BR of the Higgs boson decay to \gammad. The same exclusion limits have been derived
for the 800 GeV heavy scalar. Table 2 shows the ranges in which the \gammad~lifetime ($c\tau$) is excluded
at the 95\% CL for the \mH = 125 GeV \higgstwo+X and \higgsfour+X, assuming a BR
of 10\%. It also shows the 95\% CL $c\tau$ exclusion ranges for the \mH= 800 GeV, assuming a 5 pb
production cross section and a 100\% BR to \gammad.


\begin{table}[h]
\begin{center}
\begin{tabular}{l|c|c }  

FRVZ model & \mH [GeV] & Excluded $c\tau$ [mm]\\ \hline
\higgstwo+X & 125 & 2.2 $ \leq c\tau \leq $ 111.3\\
\higgsfour+X & 800 & 3.8 $ \leq c\tau \leq $ 163.0\\
\higgstwo+X & 125 & 0.6 $ \leq c\tau \leq $ 63.0\\
\higgsfour+X & 800 & 0.8 $ \leq c\tau \leq $ 186.0\\ 
\end{tabular}
\caption{Ranges of \gammad~lifetime ($c\tau$) excluded at 95\% CL for \higgstwo+X and \higgsfour+X
assuming for the 125 GeV Higgs a 10\% BR and the Higgs boson SM gluon fusion production cross section,
and for the 800 GeV Higgs-like scalar a \sigmaBR = 5 pb~\cite{LJ}. TYPE2-TYPE2 events are excluded.}
\label{tab:ctau}
\end{center}
\end{table}



\begin{thebibliography}{99}


\bibitem{Strass} 
  M. J. Strassler and K. M. Zurek, 
  Phys.\ Lett.\ B {\bf 651}, 374 (2007)

\bibitem{FRVZ} 
  A. Falkowski, J. T. Ruderman, T. Volansky and J. Zupan, 
  JHEP 05 077 (2010)
 
 \bibitem{LJ}
  ATLAS Collaboration,
  ATLAS-CONF-2016-042 \url {https://cds.cern.ch/record/2206083}
  
\bibitem{Aad:2012tfa} 
  ATLAS Collaboration,
  Phys.\ Lett.\ B {\bf 716}, 1 (2012)
  [arXiv:1207.7214 [hep-ex]].
  
 
  
\end{thebibliography}
\end{document}